\documentclass{aa}  
\usepackage{siunitx}

\newcommand{\G}{\,{\rm G}}

\newcommand{\nG}{\,{\rm nG}}

\newcommand{\Mpc}{\,{\rm Mpc}}
\newcommand{\cMpch}{\,h^{-1}{\rm cMpc}}

\newcommand{\Mpch}{\,h^{-1}{\rm Mpc}}

\newcommand{\cMpc}{\,{\rm cMpc}}

\newcommand{\ckpch}{\,h^{-1}{\rm ckpc}}

\def\blue{\textcolor{blue}}
\usepackage{xcolor}

\usepackage[normalem]{ulem}

\usepackage{amsmath}
\usepackage{siunitx}

\usepackage{hyperref}
\hypersetup{
    colorlinks = true,
    linkcolor = blue,
    citecolor = blue,
    urlcolor  = blue,
}

\usepackage{graphicx}
\usepackage{txfonts}

\begin{document}

   \title{Simulated rotation measure sky from primordial magnetic fields}


   \author{S. Mtchedlidze,
          \inst{1,2,3}
          F. Vazza,\inst{1,3}
          X. Du, \inst{4,5}
          E. Carretti, \inst{3}
          C. Stuardi, \inst{3}
          S. P. O'Sullivan \inst{6}
          }

   \institute{Dipartimento di Fisica e Astronomia, Universit\'{a} di Bologna, Via Gobetti 92/3, 40121, Bologna, Italy\\
              \email{salome.mtchedlidze@unibo.it}
         \and 
             School of Natural Sciences and Medicine, Ilia State University, 0194 Tbilisi, Georgia
         \and
             INAF Istituto di Radioastronomia, Via Gobetti 101, 40129 Bologna, Italy
        \and
             Department of Physics and Astronomy, University of California, Los Angeles, CA 90095, USA
        \and
             Carnegie Observatories, 813 Santa Barbara Street, Pasadena, CA 91101, USA
        \and
            Departamento de Física de la Tierra y Astrofísica \& IPARCOS-UCM, Universidad Complutense de Madrid, 28040 Madrid, Spain
             }

  \abstract{
 Primordial magnetic fields (PMFs) --- magnetic fields that originate in the early Universe and permeate the cosmological scales today --- can explain the observed micro-gauss-level magnetisation of galaxies and their clusters. In light of current and upcoming all-sky radio surveys, PMFs have drawn attention not only as major candidates for explaining the large-scale magnetisation of the Universe, but also as potential probes of early-Universe physics. While much recent work focuses on constraining the strength of PMFs, it remains challenging to constrain their structure (coherence scale). In this paper, using cosmological simulations coupled with light-cone analysis, for the first time we study the imprints of the PMF structure on the mean rotation measure (RM) originating in the intergalactic medium (IGM), $\langle \mathrm{RM_{IGM}}\rangle$. We introduce a new method for producing full-sky $\mathrm{RM_{IGM}}$ distributions. By analysing the autocorrelation of $\mathrm{RM_{IGM}}$ on small and large angular scales, we find that PMF structures show distinct signatures. The large-scale uniform model (characterised by an initially unlimited coherence scale) leads to correlations up to $\SI{90}{\degree}$, while correlations for small-scale stochastic PMF models drop by a factor of $100$ at $ 0.17, 0.13$, and $\SI{0.11}{\degree}$ angular scales, corresponding to $5.24, 4.03$, and $3.52 \Mpc$ scales (at $z=2$ redshift depths) for magnetic fields with comoving $3.49, 1.81, 1.00 \Mpch$ coherence scales, respectively; the correlation amplitude of the PMF model with comoving an $\sim 19 \Mpch$ coherence scale drops by only a factor of $10$ at $\SI{1.0}{\degree} (30.6 \Mpc)$. These results suggest that improvements in the modelling of Galactic RM will be necessary to investigate the signature of large-scale correlated PMFs. A comparison of $\langle \mathrm{RM_{IGM}}\rangle$ redshift dependence obtained from our simulations with that from the LOFAR Two-metre Sky Survey shows agreement with the our previous upper limits' estimates of the PMF strength derived from RM-rms analysis.
}
  
\titlerunning{Rotation measure sky from primordial magnetic fields}
\authorrunning{Mtchedlidze et al.}
   \maketitle

\section{Introduction}
\label{sec:Intro}

Primordial magnetic fields (PMFs), which are weak seed magnetic fields that already exist during inflation or are generated later during reheating and phase-transitions (see \citealt{Subramanian2016,Vachaspati2020} for recent reviews), have been hypothesised to explain the magnetisation of the Universe on different scales. These fields, as well as seed fields from alternative scenarios (e.g. astrophysical seed fields generated in the post-recombination Universe), need to be amplified during structure formation via the combined effect of adiabatic compression and turbulent dynamo action to explain the large-scale magnetisation of the Universe. Diffuse radio emission, visible through synchrotron radiation, on galaxy cluster scales \citep{vanWeerenetal2019}, the so-called bridges scales \citep{Botteonetal2018,Govonietal2019,Botteonetal2020,
Pignataroetal2024} 
and (galaxy-group) filaments scales \citep{Vernstrometal2021,2023SciA....9E7233V} hint at the existence of 
$\sim 1-0.01 \rm \mu G$ magnetic
fields in these environments, respectively. 
Filament-scale fields cannot be explained by only considering astrophysical sources of magnetisation (e.g. magnetised jets from active galactic nuclei or stellar winds and battery fields transported on galaxy scales; see e.g. \citealt{FurlanettoLoeb2001,NaozNarayan2013,DurriveLanger2015,Attiaetal2021,va21magcow,Garaldietal2021}), whereas PMFs can naturally pervade cosmological scales.

High-energy gamma-ray observations of TeV blazars, on the other hand, have been used to place constraints on the strength of the magnetic fields that fill the voids of the intergalactic medium (IGM); such magnetic fields, with strengths higher than 
$7.1 \times 10^{-16}\G$ 
(\citealt{Aharonianetal_2023}; 
see also \citealt{VERITAS2017,Fermi_Lat2018,Acciarietal2023}), should deviate electron-positron pairs, \blue{which} are deposited in the IGM through the interactions of TeV gamma rays with the extragalactic background light, and \blue{should} broaden the secondary GeV (cascade) emission \citep{Aharonianetal1994}. This GeV emission results from the electron-positron pairs that are upscattered with the cosmic microwave background (CMB) photons. The lack of the expected GeV bump observed in some of the blazar spectra \citep{Neronov2010,Tavecchioetal2010} strengthens the idea of primordial cosmic magnetism \citep[see also][for a recent work, and references therein]{Tjemslandetal2024}; although it has also been argued that the cooling time due to interactions between the relativistic pair beams and denser IGM can be shorter than the inverse-Compton cooling time, leading to plasma instabilities, and to a similar absence of the GeV bump (\citealt{Brodericketal2012,Brodericketal2018,Vafinetal2019}; see also \citealt{Alawashra2022,Alawashraetal2025} who argued that large-scale magnetic field can itself affect such interactions and suppress a further growth of instabilities). 
However, laboratory experiments have recently shown that the instability is suppressed unless the blazar beam is perfectly collimated or monochromatic \citep{Arrowsmithetal_2025}. Therefore, the lower limit on the magnetic field strength inferred from $\gamma$-ray observations remains robust.

Observations of extragalactic polarised radio emission and its Faraday rotation made with a new generation of radio telescopes independently hint at the magnetisation of filaments' scales \citep{Vernstrometal2019,OSullivanetal2020,2020A&A...638A..48S,Carrettietal2022,Carrettietal2025}; studying these structures is promising in the search for PMF imprints since magnetic fields in these environments  are expected to be mainly amplified via adiabatic compression \citep{Vazzaetal2014}.\footnote{We note, however, that how turbulent dynamo affects PMFs with different coherence scales is still not a trivial question. See e.g., \cite{Mtchedlidzeetal_2023}. }
The total rotation measure (RM) quantifies the rotation of the polarisation plane of a linearly polarised emission, which is caused by the birefringent foreground magnetised medium along the line of sight (LOS), from the observer to the source. It is usually expressed as
\begin{equation}
\label{eq:RM} 
\mathrm{RM} = 0.812 \int_{0}^{l} (1+z)^{-2}  \frac{n_e}{[\text{cm}^{-3}]}  \frac{(\mathbf{B} \cdot \hat{\mathbf{e}})}{[\mu G]} ~\frac{dl}{[\text{pc}]} ~~~~\frac{\text{rad}}{{\text{m}}^{2}},
\end{equation}
where $n_e, B$, and $\hat{\mathbf{e}}$ denote electron number density, magnetic field, and unit vectors, respectively; the latter  indicates the propagation direction of an emission. The total RM includes contributions from the magnetised medium of the source, $\mathrm{RM}_{\text{source}}$, of the IGM $\mathrm{RM_{IGM}}$, and of our own Galaxy, $\mathrm{RM}_{\text{Gal}}$:
\begin{equation}
\label{eq:RM-contributions}
\mathrm{RM} = \mathrm{RM}_{\text{source}} + \mathrm{RM}_{\text{Gal}} + \mathrm{RM}_{\text{IGM}}.
\end{equation}

The authors of earlier works \citep[][]{Kawabataetal1969,Fujimotoetal1971,Nelson1973a,Nelson1973b,Vallee1975,KronbergNormandin1976,Kronbergetal1977} have already made efforts to obtain the residual rotation measure (RRM), RRM = RM $- \mathrm{RM}_{\text{Gal}}$.\footnote{Hereafter when referring to RRM obtained from simulations we will usually refer to it as $\mathrm{RM_{IGM}}$, or we might interchangeably use $\mathrm{RM_{IGM}}$ and RRM.} These studies mainly focused on studying the root mean square (rms), variance and mean statistics of the RRM and their dependence on redshift, with the aim of understanding the IGM magnetic field strength and structure \citep[see also][]{Blasietal1999,Pshirkovetal2016}. For example, \citet{Nelson1973a} showed that if there is a stochastic magnetic field in the IGM, its rms should increase with redshift \citep[see also][]{AkahoriRyu2010,AkahoriRyu2011}; while \cite{Kawabataetal1969} showed that if one studies polarised sources with galactic latitudes $\geq \SI{35}{\degree}$ \citep[so that the $\mathrm{RM}_{\text{Gal}}$ is minimised]{Fujimotoetal1971}, most of the Faraday rotation takes place in the IGM; 
they showed that RM is correlated with $z \, \cos \theta$, where $z$ is the redshift of the source and $\theta$ is the angle between the direction of the source and magnetic field. 
\citealt{Kawabataetal1969} confirmed the conclusions of \citealt{Sofueetal1968} concerning the presence of a large-scale-correlated extragalactic magnetic field.
\cite{Kronberg_1977}, \cite{Vallee1990}, and \cite{Kolatt_1998} also noticed that if there is uniform (large-scale correlated) extragalactic magnetic field, then the dipole structure of the mean RM, $\langle \mathrm{RM}\rangle$, should be seen in the RM sky.

Rotation measures from the LOFAR (LOw-Frequency ARray) Two-metre
Sky Survey \citep[LoTSS; observations in the 144 MHz regime]{Shimwelletal2022,OSullivanetal2023} have been analysed in recent studies \citep{Carrettietal2022,Carrettietal2023,Carrettietal2025}. These observations have the advantage that LOFAR RM sources are only affected by filaments and voids. 
The aforementioned works have confirmed that most of the RRM, measured at low frequencies, comes from magnetic fields in the IGM {\citep[see also][for how $\mathrm{RM_{source}}$ can also be subtracted]{Vernstrometal2019,OSullivanetal2020}, and only 21 percent from magnetic fields within galaxy clusters and the circumgalactic medium \citep[see also][]{Andersonetal2024}, with RRM rms showing an increasing trend even at high redshifts ($z \sim 3$). In this latter work, the authors also used a more refined analysis for removing the RM$_{\text{Gal}}$ contribution from the total RM. It should be noted that there are still uncertainties in the analysis, both in removing the Galactic foreground contribution from the total RM, as well as in the comparison of simulated RRMs with the observation trends: for example due to simulations being unable to model the local Galactic environment and large-scale structure effects simultaneously (without the use of sub-grid physics). Nevertheless, \cite{Vazzaetal2025}, who present a new suite of cosmological MHD simulations tailored to reproduce the observed star formation history in galaxies, have shown that the magnetisation of filaments and voids from astrophysical sources alone, such as from active galactic nuclei (AGN), cannot fully account for the observed RRM-rms trends. 

In \cite{Mtchedlidzeetal2024} (hereafter Paper II), we continuously stacked cosmological boxes until $z=2$ redshift depths to study the RM-rms redshift evolution for different PMF models (with different coherence scales). We compared the obtained RM-rms trends with the ones obtained from the LoTSS survey \citep{Carrettietal2023} and constrained the strength of the PMF models. 

In this work, we take a step further and use a more realistic approach for integrating Equation~\ref{eq:RM}, in order to mimic RM analysis for large fields of view (FOVs). More specifically, our analysis aims to reproduce the full-sky mean RRM map,
while posing the following question: if future observations — such as those from the Square Kilometre Array (SKA) — allow us to map the full-sky RRM, would we be able to identify the signatures of PMFs? 
We search for potentially detectable signatures of PMFs in the mean RRM since this statistic, although often neglected in recent studies under the assumption that it should be zero on average, produces PMF-dependent patterns on the sky (even when it is fluctuating around zero), which future observations might be able to detect.

The paper is organised as follows. In Section~\ref{sec:Methods}, we describe our method of producing an $\mathrm{RM_{IGM}}$ sample. In Section~\ref{sec:results} we present our results, while Section~\ref{sec:summ} gives the conclusions and outlook.

\begin{figure}[t]\label{LCs-sketch}
    \includegraphics[width=\columnwidth]{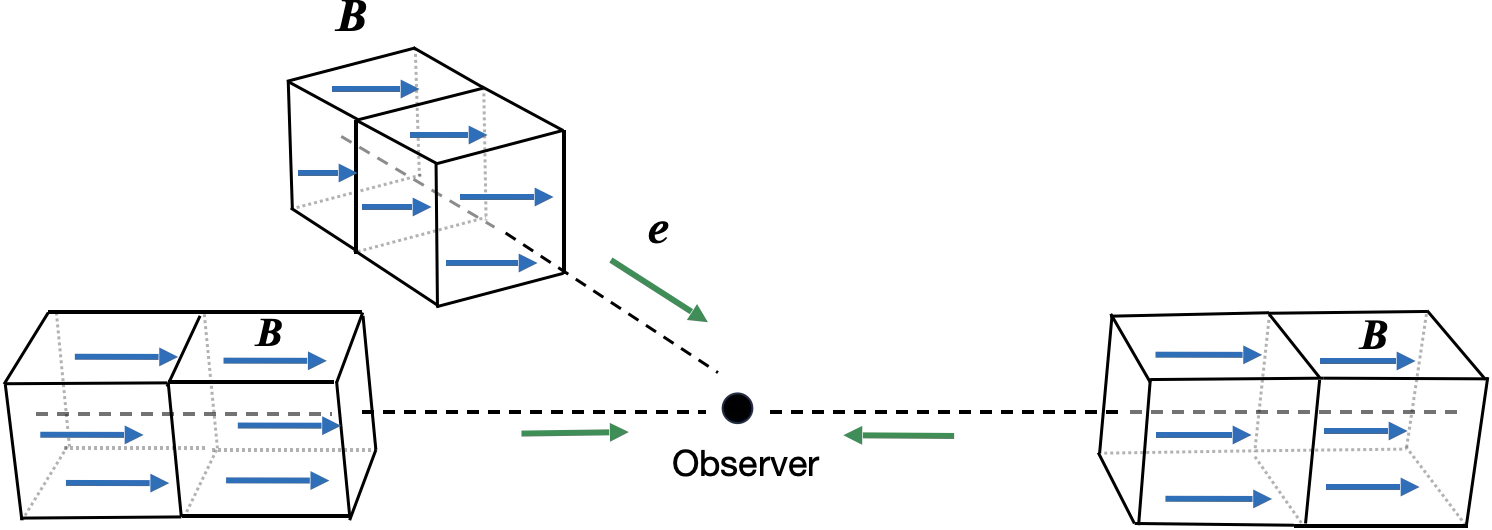}
    \caption{
    Illustration of light-cone (direction indicated with dashed lines) realisations within stacked comoving boxes. The blue and green vectors indicate the magnetic field and the $\hat{\mathbf{e}}$ unit vectors (Equation~\ref{eq:RM}), respectively.}
\end{figure}

\section{Methods}
\label{sec:Methods}

We used cosmological MHD simulations performed with the \texttt{Enzo} code \citep{Bryanetal2014}, coupled with yt \citep{yt-Turk2011} light-cone analysis \citep{Smithetal_2022} to model the IGM RM produced by PMFs. The simulation setup and light-cone methodology are described in \cite{Mtchedlidzeetal_2022} (hereafter Paper I) and in \cite{Smithetal_2022}, respectively. Here we briefly summarise that we simulated the $(135.4 h^{-1}\cMpc)^3$ (`c' referring to comoving units) comoving volume with $1024^3$ grid points, which yielded a spatial resolution of $132\,\ckpch$ and a dark matter (DM) mass resolution of $m_{\text{DM}} = 2.53\times 10^{8} M_{\odot}$. Simulations start at $z=50$ and assume a Lambda cold dark matter cosmology with the parameters $h=0.674$, $\Omega_m=0.315$, $\Omega_b=0.0493$, $\Omega_{\Lambda}=0.685$, and $\sigma_8=0.807$ \citep{Planck2018}. As described in Paper I and Paper II, in our setups fluid equations (see Equations 1-8 in \citealt{Bryanetal2014}) are solved using a second-order accurate piecewise linear method for spatial reconstruction and a second-order Runge-Kutta scheme for temporal integration (applied to both gas and DM components). Magnetic fields are evolved using the Dedner divergence-cleaning method \citep{Dedneretal2002}, and fluid discontinuities are handled with a Harten-Lax-van Leer (HLL) Riemann solver. The DM dynamics, governed by Newton's equations, is computed via an N-body approach, with particle positions interpolated to the grid. The gravitational potential is obtained by solving the Poisson equation using a fast Fourier Transform (FFT)-based solver.

We are interested in the signatures of PMFs on $\mathrm{RM}_{\text{IGM}}$; for this purpose, we studied PMFs with different coherence scales. These models are characterised by a stochastic magnetic field distribution with power spectra that peak at different scales within the simulation volume. We considered models with $3.49$ (labelled as \texttt{k25}\footnote{For small-scale stochastic PMFs, numbering in the labels indicates peak wavenumber of the power spectrum when the box size corresponds to 1 in dimensionless units.})$, 1.81$ (\texttt{k50}), and $1.00\,h^{-1}\cMpc$ (\texttt{k102}) correlation lengths and $\sim k^4$ and $k^{-5/3}$ power spectrum slopes on the left and right side of the characteristic peak, respectively, with $k$ denoting the wavenumber. We also studied the uniform magnetic field model (labelled as \texttt{u}; 
see Table 1 and Section 2.1 in Paper II for the motivation for such initial conditions), constant-strength field and a stochastic PMF model with a nearly scale-invariant spectrum ($\sim k^{-1}$; labelled as \texttt{km1}); see Figure~\ref{fig:inits_PS} and Table~\ref{table:1}.
Here we remind the reader that while the spectra of k25, k50, and k102 models resemble PMFs from phase transition magnetogenesis, their coherence scales are much larger than what is currently predicted by the theory at the recombination epoch \citep[see e.g.][and references therein]{BanerjeeJedamzik2004,Brandenburgetal2017,HoskingSchek_2023}; although it should also be mentioned that a precise picture of how PMFs evolve across recombination and until redshift $z \sim 50$ (initial redshift of our simulations) is still lacking.}\footnote{See for example \citealt{Trivedietal2018,Jedamziketal_2025,Schiff_Venumadhav_2025arXiv} where numerical and analytic efforts have been made recently to understand the evolution of PMFs and its impact on recombination.} On the other hand, our uniform and scale-invariant cases would correspond to inflationary PMF models whose coherence scales are not bound by the Hubble horizon during phase-transitions (see e.g. \cite{DurrerNeronov2013} for a review ). 
The mean magnetic field strength for all PMF models is $\sim \nG$, which is on the order of, or lower, than the limits from the \citealt{Plancketal2016} CMB analysis. CMB analysis accounts for various effects of PMFs, such as the impact of PMFs on the CMB temperature and polarisation; similar to \cite{Carrettietal2025}, \citealt{Plancketal2016} put constraints on the PMF value smoothed on megaparsec scales.

\begin{table}
\caption{Studied PMF models and their characteristics. See also Figure~\ref{fig:inits_PS} for the corresponding power spectra.}            
\label{table:1}     
\centering                          
\begin{tabular}{c c c c}        
\hline\hline                 
Simulation label & Correlation length, $\lambda_B$ & Normalisation  \\    
\hline                        
   \texttt{u} & -- & $1\nG$ \\      
   \texttt{km1} & $18.80 \cMpch $& $1\nG$   \\
   \texttt{k25} & $3.49 \cMpch $& $1\nG$     \\
   \texttt{k50} & $1.81 \cMpch $& $1\nG$   \\
   \texttt{k102}& $1.00 \cMpch $& $1\nG$  \\ 
\hline                                   
\end{tabular}
\end{table}

The stochastic PMF initial conditions are generated as Gaussian random fields using the \textsc{Pencil Code} \citep{JOSS} initial conditions’ routine. We note that these initial conditions are not self-consistently coupled with cosmological initial conditions generated by the \texttt{Enzo} code; i.e. matter and velocity perturbations are generated according to the standard $\Lambda$CDM prescriptions, not accounting for the initial magnetic field fluctuations. Nevertheless, as studied in \cite{Kahniashvilietal2013}, \cite{Sanatietal2020}, \cite{Katzetal2021}, and \cite{Pavicevicetal_2024}, PMF effects on the matter power spectrum are expected on galaxy scales;
therefore, we do not expect that simulated $\mathrm{RM}_{\text{IGM}}$ evolution will be significantly altered by such processes (see also \citealt{Adietal2023}, \citealt{Cruzetal_2024}, and \citealt{Ralegankaretal2025}).

The simulated cosmological boxes were stacked up to redshift $z=2$ to construct light cones with a \SI{2}{\degree} FOV and \SI{20}{\arcsecond}
image resolution, and were then used to generate RM maps out to $z\lesssim 2$ redshift depths. This procedure follows the same methodology as in Paper II, including a filtering technique with which we filter out all high-density regions (we keep $\rho / \langle \rho \rangle < 1.3\times 10^{2}$ regions, with $\rho$ being the gas density) to focus only on the IGM environment when integrating Equation~\ref{eq:RM}; however, unlike the previous work, in this study we produced each light-cone realisation by varying the direction of the unit vector $\hat{\mathbf{e}}$ in Equation~\ref{eq:RM}. 
The purpose of this simple operation is to allow us to `observe' the same simulated Universe in RM from different viewing angles; that is, we can produce a realistic full-sky distribution of $\mathrm{RM_{IGM}}$ around the fixed observer by effectively rotating the viewing angle with respect to the same simulated volume.

We made our calculations computationally less expensive by first producing 100 light-cone maps of the $\mathrm{RM}_{\mathrm{x}}, \mathrm{RM}_{\mathrm{y}}, \mathrm{RM}_{\mathrm{z}}$ fields; using yt we integrated
\begin{equation}
\label{eq:RMz} 
\mathrm{RM_{\text{s}}} = 0.812 \int_{0}^{l} (1+z)^{-2}  \frac{n_e}{[\text{cm}^{-3}]}  \frac{B_s}{[\mu G]} \frac{dl}{[\text{pc}]} ~~\frac{\text{rad}}{{\text{m}}^{2}},
\end{equation}
where $s=\mathrm{x},y,z$, $dl=d\mathrm{x}$ for the uniform model, and $dl$ is randomly directed either along $\mathrm{x}$, $y$, or $z$ for stochastic PMF cases for each simulated snapshot used in the light-cone stack.
We used these initial 100 light-cone maps to generate $10^4$ light-cone realisations 
by multiplying the randomly chosen 
$\mathbf{RM}_{\mathrm{initial}} \equiv (\mathrm{RM}_{\mathrm{x}}, \mathrm{RM}_{\mathrm{y}}, \mathrm{RM}_{\mathrm{z}})$ 
light-cone map with a
random 
$\hat{\mathbf{e}} \equiv (\mathrm{e_x,e_y,e_z})$
vector.
That is, while randomly choosing $\mathbf{RM}_{\mathrm{initial}}$ from the original sample, we also generated $10^4$ unit vectors (see Equation 4 below) to mimic different light-cone directions; such a method does not require light-cone projections for each line of sight direction around observer.
This procedure is mathematically equivalent to the procedure where different $\hat{\mathbf{e}}$ are applied directly to the stacked comoving boxes during the integration of the RM field within yt (Equation~\ref{eq:RM}); see Figure~\ref{LCs-sketch}, which depicts two such light-cone realisations with different $\hat{\mathbf{e}}$ vectors applied to stacked comoving boxes and with a simulated uniform magnetic field.
Here we also note that the light-cone projection module in yt only performs integrations along the coordinates axes; it does not do off-axis projections. This is effectively equivalent to keeping the magnetic-field direction fixed while rotating the simulation box when filling up the cosmic volume. We emphasize that, for the purpose of this work, the key factor is the angle between the magnetic field and the LOS direction. We will explore alternative approaches, such as off-axis projections, in future work.

It can be shown that the angle-dependence of correlation of full-sky RM maps obtained with the aforementioned procedure is proportional to $\cos(\Delta)$, where $\Delta$ is an angle between 
the $\hat{\mathbf{e}}_i$ and $\hat{\mathbf{e}}_j$ unit vectors (where $i,j$ runs over all chosen unit vector pairs; see Appendix~\ref{app:RMangle}).  
Therefore, multiplying by unit vectors introduces a correlation in the large-angle ($\gtrsim \SI{2}{\degree}$) correlation function if $\langle \mathbf{RM}_{\mathrm{initial}} \rangle \neq 0$. In the uniform magnetic field case, this behaviour is expected \citep[see e.g.][]{Kronberg_1977,Vallee1990,Kolatt_1998} and should also be revealed by our analysis. In stochastic cases, $\langle \mathbf{RM}_{\mathrm{initial}}\rangle$ is expected to be zero, considering that the magnetic field has no preferred direction. However, in practice, due to the finite box size, $\langle \mathbf{RM}_{\mathrm{initial}} \rangle$ obtained from the light-cone projection has a non-zero residual, leading to artificial large-angle correlation. To eliminate this artificial correlation in these cases, we further multiplied the generated RM realisations by the rotation matrix, $R$ (randomly varying for each generated realisation), so that our final RM realisation is calculated using the equation,\footnote{For the uniform case only, we set $R=I$.}
\begin{equation}
\label{eq:RM_final}
    \mathrm{RM_{final}} =  \mathrm{\hat{\mathbf{e}}} \cdot ( R~ \mathrm{\mathbf{RM}_{initial} ) }.
\end{equation}
For each realisation the direction of $\hat{\mathbf{e}}$ was randomly selected 
by sampling $\phi$ and $\cos \theta$ from uniform distributions over the $[0,2\pi]$ and $[1,-1]$ ranges, respectively. Here $\phi$ and $\theta$ are the polar and azimuthal angles in spherical coordinates.
Using this computationally efficient method (than the standard approach of directly integrating Equation~\ref{eq:RM}) enables us to produce a large sample of $\mathrm{RM_{IGM}}$ realisations.

\section{Results}
\label{sec:results}

\begin{figure}[t]
    \includegraphics[width=\columnwidth]{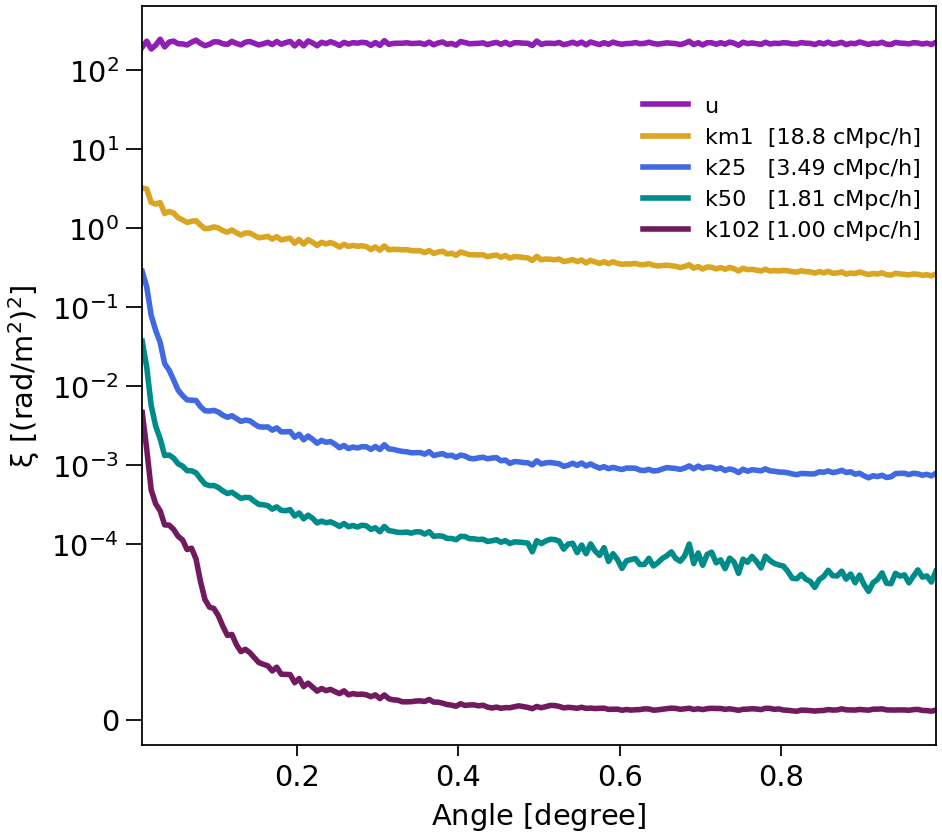}
    \caption{Average of $\mathrm{RM_{IGM}}$ autocorrelation functions obtained from $10^4$ light-cone realisations for each model at $z=2$ redshift depth; coherence scales of the PMF models are also indicated in the legend.
    }   
    \label{fig:cross_corr_2def}
\end{figure}

We start presenting our results by analysing the 1D autocorrelation \citep[see][for similar analysis]{GhellerVazza2020}, $\xi$, of $\mathrm{RM_{IGM}}$ within a \SI{2}{\degree} FOV first. We calculated the 2D correlation, $C$, of RM maps (for convenience, in the equation below, we removed subscript IGM for RM)
using \texttt{Scipy} \citep[][]{2020SciPy-NMeth},
\begin{equation}
    C[k,l] = \sum_{i,j}^{N-1, M-1} \mathrm{RM_{ij}}~\mathrm{RM_{i-k,j-l}},
\end{equation}
where $k$ runs from $-(N-1)$ to $(N-1)$, and $l$ runs from $-(M-1)$ to $(M-1)$, and $N, M$ are lengths of 2D $\mathrm{RM}$  arrays in the first and second dimensions, respectively. To calculate $\xi$, we then binned $C$ in 1D radial bins, with a maximum radius corresponding to half of our FOV at $z=2$ redshift depth.

In Figure~\ref{fig:cross_corr_2def} we show the correlation function averaged over all $10^4$ realisations for all PMF models. 
We see from the figure that the amplitude of the correlation function varies by orders of magnitude across different PMF models. As we showed in Paper I and Paper II, the $\mathrm{RM}_{\text{IGM}}$ amplitude depends on the coherence scale of the magnetic field; fields with larger coherence scales lead to larger magnetic fields in the IGM and, therefore, larger RMs in this environment. 
Since $\xi(\SI{0}{\degree})= (\mathrm{RM}_\text{rms})^2$, the first point of the correlation function is larger for our u and km1 models, which have the largest coherence scales. We also see that $\xi$ is flat for the uniform model while it decreases for the stochastic fields (models: km1, k25, k50, and k102). 
The flatness of the function in the uniform case reflects the fact that the structure of this model undergoes insignificant changes on large scales. 
In the case of k25, k50, and k102, $\xi$ is further characterised by a sharp decrease as angles increase up to $\sim \SI{0.2}{\degree}$;
angular scales at which the amplitude of the correlation function drops by a factor of $10^2$ are 
$\SI{0.17}{\degree} ~(5.24 \Mpc), \SI{0.13}{\degree} ~(4.03 \Mpc)$, and $\SI{0.11}{\degree} ~(3.52 \Mpc$; $z=2$ redshift depth), respectively; in the case of km1 model, correlation drops by only a factor of 10 at $\SI{1.0}{\degree}~ (30.6 \Mpc)$.
Thus, the correlation decreases faster for small-scale-correlated fields as FOVs increase in the RM sky. We also checked correlation for the absolute value of $\mathrm{RM_{IGM}}$, which does not show the aforementioned features and remains flat at all radii for all PMF models (with differences in the amplitude of the autocorrelation).

In Paper II, we found that small-scale PMFs (k25, k50, k102) show degeneracy in their RM-rms redshift-dependence trends (the shapes of RM-rms redshift evolution are similar for the aforementioned models).
The distinctive features of $\xi$ (such as declining trends) found for the stochastic cases in the $\langle \mathrm{RM_{IGM}} \rangle$ analysis are therefore crucial for constraining the coherence scale of such PMF models; 
to the best of our knowledge, the work of \citet{Kolatt_1998} was the first to argue that correlation analysis of RM is more beneficial for understanding the structure of PMFs.
The density of RM points in the current LOFAR RM grid is $\sim 0.4$ points per square degree, which is approximately seven orders of magnitude smaller than the density of our $\mathrm{RM_{IGM}}$ maps. The RM grid obtained with SKA observations will have $\gtrsim 100$ polarised sources per $\textrm{deg}^2$. In our future work, we will estimate whether uncertainties in the calculated $\xi$ values --- when accounting for the same grid densities as in SKA observations --- still allow us to distinguish small-scale-correlated PMFs.
We also note that RMs obtained through future SKA observations might still have larger uncertainties than required for distinguishing these small-scale, $\nG$-strength PMFs;
however, the amplitude of $\xi$ will be higher for larger initial normalisations of the same PMF models, and therefore future data will be able to start constraining the strength of such models. 

\begin{figure*}[t]\label{fig:SkyMaps}
    \includegraphics[width=\textwidth]{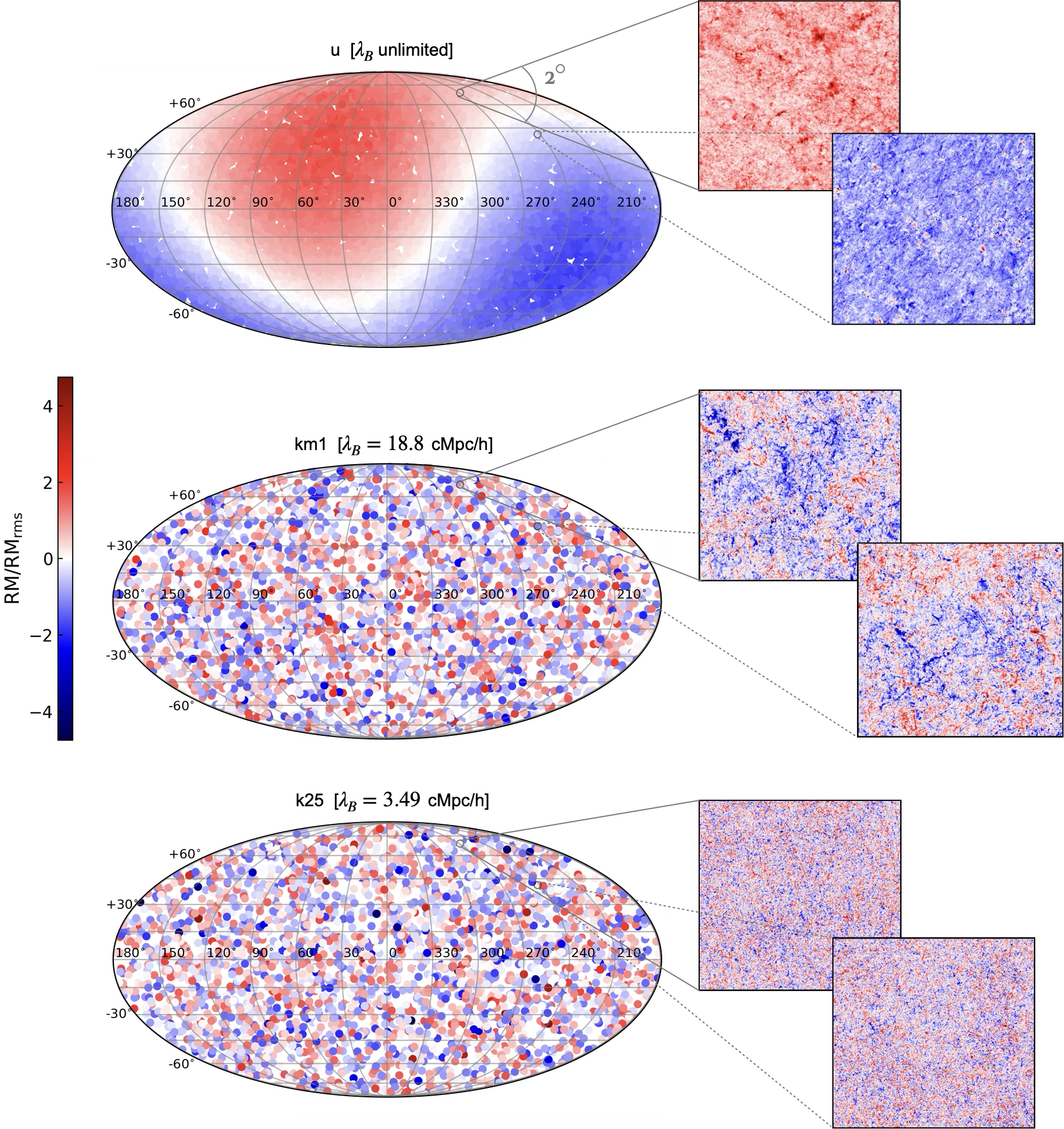}
    \caption{
    Constructed $\mathrm{RM}_{\text{IGM}}$ maps [rad/m$^2$] for three PMF models studied in this work at $z=2$ redshift depth. The left part of the figure depicts the mean $\mathrm{\langle RM_{\text{IGM}} \rangle}$ distribution in Galactic coordinates with 10000 and $\sim 3300$ light-cone realisations in the uniform, and km1 and k25 cases, respectively (in the latter cases a smaller number of realisations was chosen for a better visualisation). The right part shows example 2D light-cone maps (with $\SI{2}{\degree}$ FOVs and \SI{20}{\arcsecond} image resolution) used for calculating $\mathrm{RM_{IGM}}$ statistics. 
    }   
\end{figure*}
The correlation observed in the uniform model in Figure~\ref{fig:cross_corr_2def} is then expected to extend even on larger scales. The $\mathrm{\langle RM_{IGM} \rangle}$ full-sky map, extracted from different $\theta$- and $\phi$-dependent realisations as a mean from each $\mathrm{RM_{IGM}}$ 2D map at $z=2$, is shown in Figure~\ref{fig:SkyMaps} in Galactic coordinates; as an example we also show 2D maps from two realisations. The figure illustrates the large-scale correlations for the uniform case, as predicted by previous work \citep[see e.g.][]{Kronberg_1977,Vallee1990,Kolatt_1998}. A dipole structure of the mean $\mathrm{RM_{IGM}}$ observed across the whole sky in the uniform PMF case indicates that the magnetic field, which is chosen to be along the diagonal in our simulations, produces positive and negative RMs around the observer (a dipole structure would be present regardless of the orientation of the uniform magnetic field). This further shows that if the PMF coherence scales is larger than the Hubble horizon, its imprints must be revealed in all-sky $\mathrm{RM_{IGM}}$ observations. For the stochastic models, as mentioned earlier, and as we could already see by analysing Figure~\ref{fig:cross_corr_2def}, we do not expect large-angle correlations since their coherence scales are smaller than the simulated boxes, and thus no such structures should be seen in the RM sky. On the other hand, 2D maps (right part of Figure~\ref{fig:SkyMaps}) confirm the trends observed in Figure~\ref{fig:cross_corr_2def}; in particular, we see larger correlated structures in the km1 case compared to the structures seen for the k25 model.
\begin{figure}[t]
    \includegraphics[width=\columnwidth]{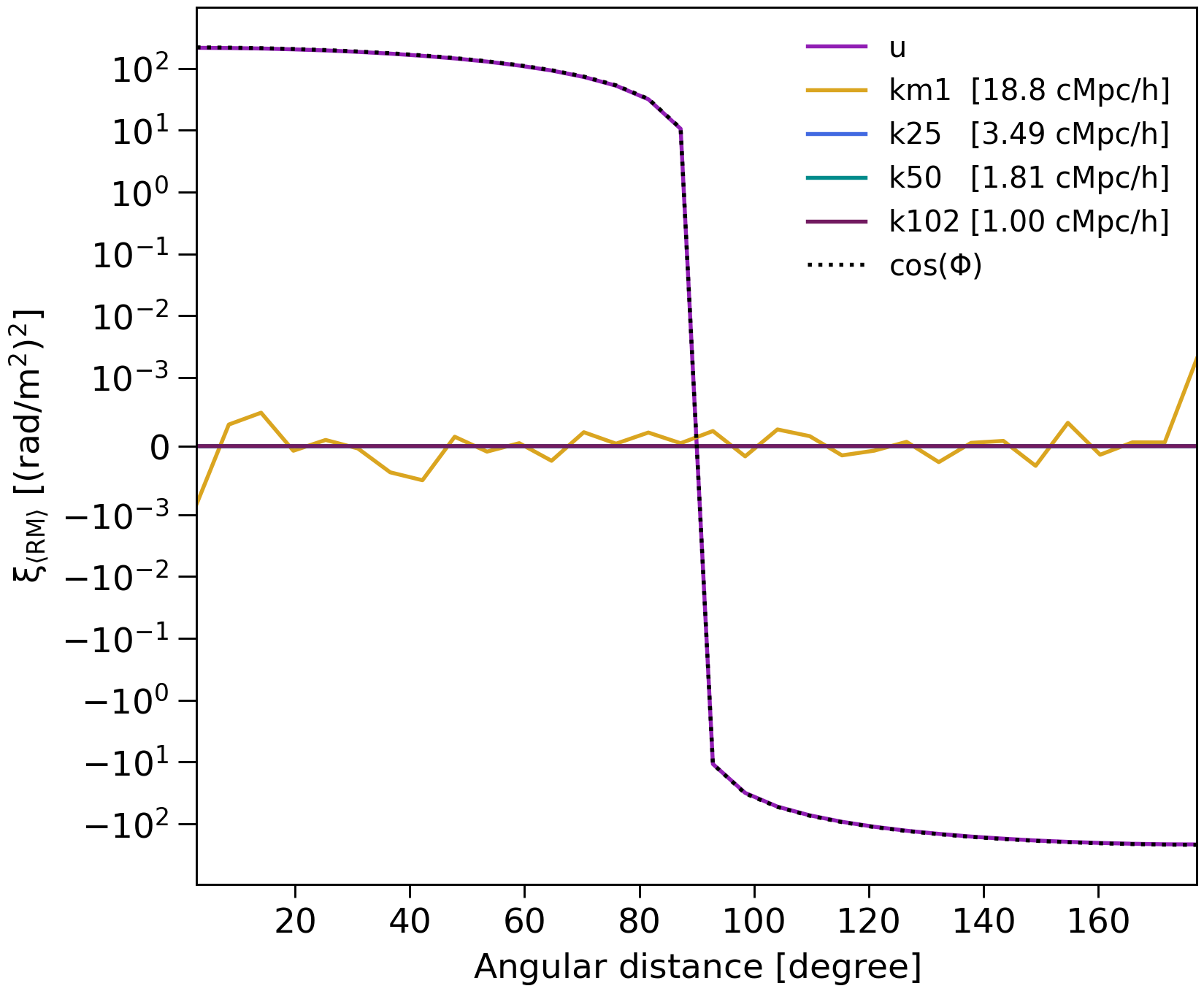}
    \caption{
    Angular autocorrelation of full-sky mean $\mathrm{RM_{IGM}}$ (Figure~\ref{fig:SkyMaps}) for the uniform and stochastic PMF models (with coherence scales indicated in the legend). The cosine function is also shown for reference, with $\Phi$ indicating the angular separation.  
    }   
\label{CrossCorrLargeAngle}
\end{figure}

The mean $\mathrm{RM_{IGM}}$ full-sky maps are quantitatively analysed in Figure~\ref{CrossCorrLargeAngle}. The correlation function for these maps is calculated by first finding all unique pairs in the $\langle \mathrm{RM_{IGM}} \rangle$ sample and their angular distances, which are then profiled in Figure~\ref{CrossCorrLargeAngle}.
As was already expected from analysing such maps, the correlation function for all models fluctuates around zero; 
although in the uniform case, it shows positive and negative correlations on the order of $10^2 \mathrm{(rad/m^2)^2}$ for angles $\lesssim \SI{90}{\degree}$ and for angles larger than $\SI{90}{\degree}$, respectively; the correlation function shape reproduces cosine function, as predicted (see Appendix \ref{app:RMangle}). These angular scales correspond approximately gigaparsec ($z=2$ redshift depths) scales;
therefore, we expect that even with low RM grid densities, the current LoTSS survey should be able to 
detect traces of the PMF with coherence scales larger than the Hubble horizon, provided that the accuracy in the determination of RMs (also including the removal of foreground contamination) is similar to that of simulations. We defer a more detailed study of this question to future work, 
while emphasising here that the detection 
of such large-scale correlated PMFs will also depend on $\mathrm{RM_{Gal}}$ removal techniques; 
in other words, the current reconstruction of $\mathrm{RM_{Gal}}$ relies on an assumption that there should not be any large-angle ($\gtrsim \SI{0.02}{\degree}$) correlations in the sky that come from the extra-Galactic processes \citep{Hutschenreuteretal2022}.
On the other hand, our $\mathrm{RM_{IGM}}$ models at high latitudes, together with $\mathrm{RM_{Gal}}$ models, can also be used to infer the parameter space in which $\mathrm{RM_{IGM}}$ can be distinguished from $\mathrm{RM_{Gal}}$ (see e.g. \citealt{Akahorietal_2013}).
\begin{figure}[htbp]
    \includegraphics[width=\columnwidth]{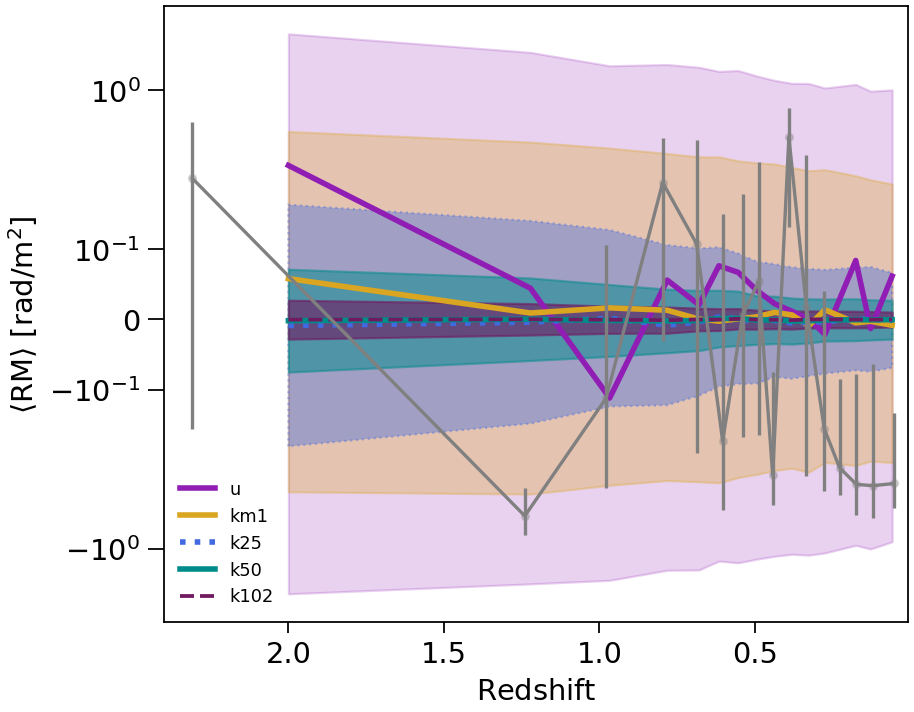}
    \caption{
    Mean $\mathrm{RM_{IGM}}$ trends of PMFs and mean RRM from LoTSS data (grey lines); the error for the latter is calculated with bootstrapping, while standard deviations are shown for the simulated RMs (lower-opacity filled colour lines). 
    }   
\label{fig:RMmeanEv}
\end{figure}

In Figure~\ref{fig:RMmeanEv} we make use of our large sample of $\mathrm{RM_{IGM}}$ maps
to give an example comparison of the simulated $\mathrm{RM_{IGM}}$ and its observed redshift evolution trends. \citealt{Carrettietal2025} used refined analysis for $\mathrm{RRM}$ rms to confirm its IGM and filamentary origin. Here, instead of rms, with grey lines, we show the results of such an analysis for the redshift evolution of mean RRM. 
These means of  $\mathrm{RM_{IGM}}$ (RRM) have been calculated from the LoTSS dataset \citep[][]{OSullivanetal2023} with 40 points in each redshift bin. For comparison purposes, in our sample we only used those  $\mathrm{RM_{IGM}}$ redshift-depth data that are close to redshift bins used in LoTSS-data analysis. 
We then randomly drew 40 $ \mathrm{RM_{IGM}}$ points (pixels) from randomly-chosen simulated 2D maps  
(1000 times) and calculated means for each subsampling and standard deviations for these means 
at each redshift depth, shown in Figure~\ref{fig:RMmeanEv}.

As we see in Figure~\ref{fig:RMmeanEv}, large-scale-correlated PMFs, such as the u and km1 models, lead to larger RM amplitudes compared to trends from small-scale stochastic models (k25, k50, k102), although all models fluctuate around zero, similarly to the data from LoTSS.  We also see that $\nG$-normalisation u and km1 cases better reproduce LoTSS mean-RM trends, while the error in the uniform case is larger and the error in the km1 model remains within the observed $\langle \mathrm{RRM} \rangle$ uncertainties. It is interesting to note that the latter model also best matches the LoTSS RM-rms trends \citep[see Figure~4 in ][]{Mtchedlidzeetal2024}. Therefore, PMFs with coherence scales $\gtrsim 20 \cMpch$ might be interesting to investigate further, with a careful rethinking how future $\mathrm{RM_{Gal}}$ reconstruction algorithms should account for the existence of extragalactic signal.
We also note that upper limits on the PMFs' strengths obtained in Paper II seem consistent with the mean RM analysis; that is, larger ($>\nG$) field strengths are allowed for small-scale-correlated PMF models and smaller field strengths ($<\nG$) for the u and km1 cases (see also \citealt{Neronovetal2024arXiv} where recent upper limits obtained from RM measurements are summarised).

As already emphasised in the Introduction, currently uncertainties in the predictions of $\langle \mathrm{RM_{IGM}} \rangle$  exist both in observations (e.g. removing foreground Galactic magnetic field contribution, or the effects from foreground radio galaxies; the latter being also relevant for simulations) and simulations. As \citealt{Vazzaetal2025} found radio galaxies also contribute to the RM-rms measurements at low redshifts \cite[see also][for a recent review]{CarrettiVazza2025}. Therefore, even if our uniform and km1 models with $\nG$ strengths better reproduce LoTSS mean-RM trends, our simulations lack the contribution from radio galaxies, the effect that we need to account for in our future studies.  However, overall, our results seem to suggest that if both simulations and observations are improved mean RM evolution can be used (probably together with RM-rms statistics) not only for constraining the amplitude of the PMF, but also its structure (coherence scale), which is the most important question in the search for the origin of cosmic magnetism.

\section{Summary and conclusions}
\label{sec:summ}
In this paper, for the first time, we explored the possibility of using $\langle \mathrm{RM_{IGM}} \rangle$ to constrain the PMF structure. Simulating PMFs with different coherence scales, analysing simulation results with light cones, and introducing a new method for obtaining the large RM sample allowed us to produce the first full-sky RM$_{\mathrm{IGM}}$ maps for different PMF models. Using this new sampling method, we produced $10^4$ realisations of RM from the original 100 light-cone maps for $z=2$ redshift depths by randomly varying angles between the magnetic field and light propagation unit vectors for each subsample, and by randomly choosing rotation matrices for stochastic PMF models (to break the artificial correlation introduced by the multiplication of unit vectors on large scales). Our results are summarised as follows:

\begin{itemize}
\item[1.] Primordial magnetic field coherence scales leave observational imprints on small- and large-angle correlation functions of $\mathrm{RM_{IGM}}$. Small-angle ($\sim \SI{1}{\degree}$) autocorrelation features can be used to distinguish different PMF models. The amplitude of correlations drop by a factor of 100 at $0.17, 0.13$, and $\SI{0.11}{\degree}$, corresponding to $5.24, 4.03$, and $3.52 \Mpc$ scales for the PMF models with $3.49, 1.81, 1.00 \cMpch$ coherence scales, respectively. The correlation amplitude for the stochastic PMF model characterised by the $18.80 \cMpch$ coherence scale decreases by a factor of 10 at $\SI{1.0}{\degree}$, corresponding to $30.6 \Mpc$. The extreme case of the uniform model, on the other hand, shows a constant amplitude of correlations within this small-angle, which extends to larger scales -- visible in the full-sky maps and large-angle correlation function. 

\item[2.] The full-sky $\mathrm{RM_{IGM}}$ maps reveal a dipole structure in the uniform case, as expected; RMs with positive and negative values extend to $\SI{90}{\degree}$, while all the other PMF models show stochastic distributions.
 
\item[3.] The large-angle correlation analysis further confirms the absence of large-scale ($\gtrsim 30 \Mpc$) correlations in these stochastic PMF cases, and existence of such correlations in the uniform case — the PMF model, which we think of as the magnetic field with an unlimited correlation length (thus, its correlation length extends to scales larger than the Hubble horizon).

\item[4.] A comparison of the produced mock  $\langle \mathrm{RM_{IGM}} \rangle (z)$ dependence with its trends from LoTSS observations shows consistency with the PMFs' upper limits trends obtained in \cite{Mtchedlidzeetal2024}, where RM rms statistics are analysed. While the PMF model with an $18.80 \cMpch$ coherence scale and $\nG$-strength shows a better agreement with the LoTSS-data, the uniform model with $B<1\nG$ and smaller-coherence scale ($\sim 3.49, 1.81, 1.00 \cMpch$) fields with $B>1\nG$ cannot be excluded solely based on the statistics of $\langle \text{RM}_{\rm IGM} \rangle$.

\end{itemize}

This  study has shown 
that there are potentially detectable signatures of PMFs on the mean $\mathrm{RM_{IGM}}$ autocorrelation function; contrary to RM-rms,  whose redshift evolution trends have degeneracies with respect to coherence scales of small-scale ($\lambda_B \sim$ $\cMpch$) PMFs, mean RM can be used to constrain the structure of the magnetic field.
Even though a constant-strength field, such as our uniform case, might be an unrealistic PMF model \citep[see, however,][]{Mukohyama2016}, it illustrates what to expect for future all-sky surveys for PMFs with very large ($\gg \rm 100 ~ Mpc$) correlation lengths; our results also require a rethinking of the $\mathrm{RM_{Gal}}$ modelling method, which currently removes extragalactic signals on large scales.
We anticipate that, for example, if the future $\mathrm{RM_{IGM}}$ all-sky maps (with an updated modelling method of $\mathrm{RM_{Gal}}$) obtained for a certain redshift  depth lack correlations visible for our uniform model, then this will constrain the PMF coherence scale to be much less than the distance to the analysed redshift depth. Then one might need to analyse the RRM data within small FOVs to understand which structure of the PMF produces observations best. Therefore, future all-sky surveys and deep surveys might be able to detect the signatures of large- and small-scale correlated PMFs, respectively. For future all-sky surveys, with larger RM-grid densities, it might also be possible to expand the RM data with spherical harmonics (similar to what has been done e.g. for CMB birefringence studies; \citealt{KosowskyLoeb1996,ScannapiecoFerreira_1997,HarariZaldarriaga_1997,Scoccolaetal_2004,Kosowsky:2004zh}) and, together with larger simulations, search for the PMF-structure imprints.

\begin{acknowledgements}
The authors acknowledge fruitful discussions with David Alonso-L\'opez, Annalisa Bonafede, Gabriella Di Gennaro, Tina Kahniashvili, Chris Riseley and David Vall\'es P\'erez. SM is supported through young scientist grant funded by the Shota Rustaveli National Science Foundation of Georgia (grant number: YS 24-758). S.M.\ and F.V.\ was supported by Fondazione Cariplo and Fondazione CDP, through grant number Rif: 2022-2088 CUP J33C22004310003 for ``BREAKTHRU'' project. 
SPO acknowledges support from the Comunidad de Madrid Atracción de Talento program via grant 2022-T1/TIC-23797, and grant PID2023-146372OB-I00 funded by MICIU/AEI/10.13039/501100011033 and by ERDF, EU.
The presented work made use of computational resources on Norddeutscher Verbund f\"ur Hoch- und H\"ochstleistungsrechnen (Germany) and Cineca (Italy; ``IsB30\_MAJIC'').

\textbf{Software and data availability.} The \texttt{Enzo} (http://enzo-project.org),
and \texttt{yt\_astro\_analysis} \citep{Smithetal_2022} codes (extension of the yt analysis toolkit \citep{yt-Turk2011} has specifically been used) used in this study are freely available. The derived data supporting the findings of this study are freely available upon request. The data for producing the figures of this paper is available at doi: \url{https://doi.org/10.5281/zenodo.18495390}.

\end{acknowledgements}

\bibliographystyle{aa}
\bibliography{PMFsRMsky}

\begin{appendix} 

\section{PMF Initial conditions}
\label{app:Inits}
%
\begin{figure}[htbp]
    \includegraphics[width=\columnwidth]{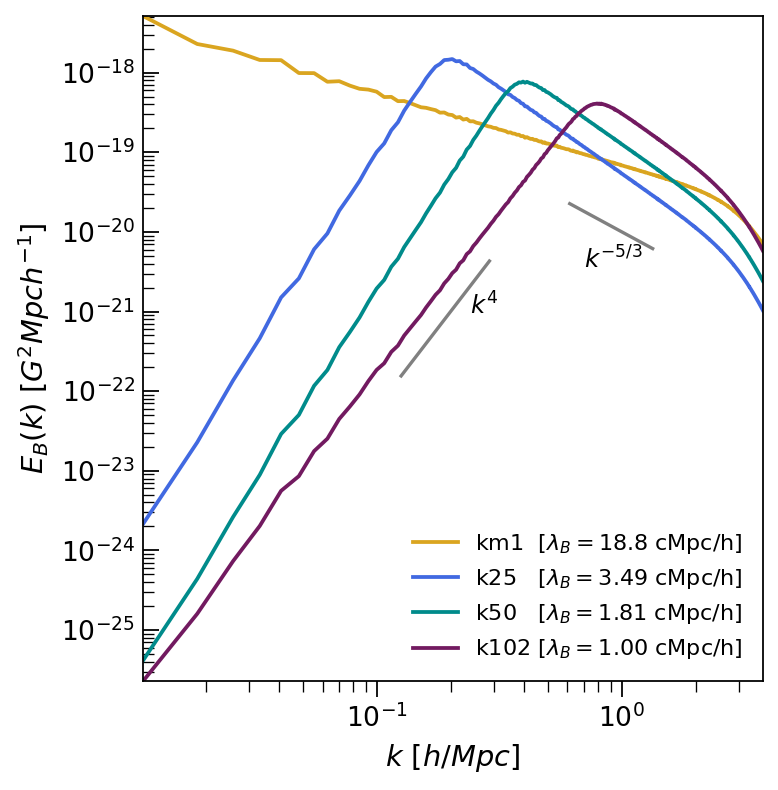}
    \caption{Initial power spectrum of the PMF models studied in this work. Adopted from \cite{Mtchedlidzeetal2024}.
    }   
\label{fig:inits_PS}
\end{figure}
In Figure \ref{fig:inits_PS} we show initial power spectrum of the PMF models for completeness. As mentioned in the main text, k25, k50, k102 cases are characterised by the turbulent spectrum at small scales (large wavenumbers), while the slope of the power spectrum in the km1 case is $\propto {-1}$ (nearly scale-invariant). For small-scale stochastic PMF models (k25, k50, k102) simulation labelling corresponds to peak wavenumbers used in the Pencil code.

\section{Introducing large-angle correlation in all-sky RM}
\label{app:RMangle}
Due to the finite box size, correlations on scales large than the simulation volume cannot be captured. For magnetic models with large intrinsic coherent length, this limitation becomes problematic when generating all-sky RM maps. To address this issue, we introduce a new method for generating light-cone projections in Sec.~\ref{sec:Methods} by defining a vector field $\mathbf{RM} \equiv (\mathrm{RM}_x, \mathrm{RM}_y, \mathrm{RM}_z)$ (see Eq.~\ref{eq:RM_final}). The RM map is then obtained by multiplying this vector field by a unit vector along the line of sight, $\mathrm{RM}=\mathbf{RM} \cdot \mathbf{e}$. In this appendix, we show that this approach naturally introduces large-angle correlations, and that the resulting angular correlation function is proportional to $\cos \Delta$, where $\Delta$ is the angular separation between two directions on the sky.

The angular correlation function of RM is given by
\begin{equation}
\xi (\Delta)  = \langle ~(\mathbf{RM}_1 \cdot \mathbf{e}_1) (\mathbf{RM}_2 \cdot \mathbf{e}_2)~\rangle,
\label{eq:cross_corr0} 
\end{equation}
where the average is taken over all pairs of $\mathbf{e}_1$ and $\mathbf{e}_2$ that satisfy $\mathbf{e}_1 \cdot \mathbf{e}_2 = \cos \Delta$. To compute Equation~\ref{eq:cross_corr0}, we first fix $\mathbf{e}_1$ and average over $\mathbf{e}_2$. Taking into account that $\langle \mathbf{e}_2 \rangle_{\mathbf{e}_2} = \cos \Delta \, \mathbf{e}_1$~\footnote{The part perpendicular to $\mathbf{e}_1$ cancels out.}, and that $\mathbf{RM}$ is independent of $\mathbf{e}$, we obtain
\begin{equation}
\xi (\Delta)  = \langle ~(\mathbf{RM}_1 \cdot \mathbf{e}_1) (\mathbf{RM}_2 \cdot \mathbf{e}_1)~\rangle \cos \Delta.
\label{eq:cross_corr1} 
\end{equation}
Here, we are left with taking the average over $\mathbf{e}_1$. We then decompose $\mathbf{RM}$ as $\mathbf{RM} = \overline{\mathbf{RM}} + \delta \mathbf{RM}$, where $\overline{\mathbf{RM}}$ is the mean vector field and $\delta \mathbf{RM}$ is the random flocculation. Substituting this expression into Equation~\ref{eq:cross_corr1} yields
\begin{equation}
\begin{aligned}
\xi (\Delta)  = & \left[ \langle ~(\overline{\mathbf{RM}} \cdot \mathbf{e}_1)^2~\rangle + \langle ~(\delta \mathbf{RM}_1 \cdot \mathbf{e}_1) (\delta \mathbf{RM}_2 \cdot \mathbf{e}_1)~\rangle \right] \cos \Delta \\
              = & \left[ \frac{1}{3} \, \overline{\mathbf{RM}}^2 + \langle ~(\delta \mathbf{RM}_1 \cdot \mathbf{e}_1) (\delta \mathbf{RM}_2 \cdot \mathbf{e}_1)~\rangle \right] \cos \Delta.
\end{aligned}
\label{eq:cross_corr2} 
\end{equation}
If we further assume that different light-cone realizations are statistically independent, for $\Delta > 0$ ($\mathbf{RM}_1 \neq \mathbf{RM}_2$) the second term on the right-hand side of Equation~\ref{eq:cross_corr2} vanishes and we get
\begin{equation}
\xi (\Delta)  = \frac{1}{3} \, \overline{\mathbf{RM}}^2 \cos \Delta.
\label{eq:cross_corr3} 
\end{equation}

\end{appendix}

\end{document}